\documentclass[onecolumn,useAMS,usenatbib]{mn2e}

\usepackage{graphicx}
\usepackage{subfigure}
\usepackage{xcolor}
\usepackage{mathtext,bbm,amsmath,amsfonts,amssymb,indentfirst,syntonly,graphicx}
\usepackage{mathtools}
\usepackage{slashbox}
\usepackage[english]{babel}
\usepackage{calc}
\usepackage{tikz}

\def\bc{\begin{center}}
\def\ec{\end{center}}
\def\be{\begin{eqnarray}}
\def\ee{\end{eqnarray}}

\title[The change of GRB polarization angles in the magnetic-dominated jet model]{The change of GRB polarization angles in the magnetic-dominated jet model}
\author[Z. Chang and H.-N. Lin]
        {Zhe Chang$^{1,2}$ and Hai-Nan Lin$^{1}$\thanks{E-mail: linhn@ihep.ac.cn.}\\
$^{1}$Institute of High Energy Physics, Chinese Academy of Sciences, 100049 Beijing, China\\
$^{2}$Theoretical Physics Center for Science Facilities, Chinese Academy of Sciences, 100049 Beijing, China}
\begin{document}

\date{Accepted xxxx; Received xxxx; in original form xxxx}

\pagerange{\pageref{firstpage}--\pageref{lastpage}} \pubyear{2014}

\maketitle

\label{firstpage}

\begin{abstract}
The polarimetric measurement on the prompt phase of GRB 100826A shows that the polarization angle changes $\sim 90^{\circ}$ between two adjacent time intervals. We will show that this phenomenon can be naturally interpreted in the framework of the magnetic-dominated jet (MDJ) model. The MDJ model suggests that the bulk Lorentz factor of the outflow increases as $\Gamma\propto r^{1/3}$, until reaching a saturated value $\Gamma_{\rm sat}$. Electrons move in the globally ordered magnetic field advected by the jet from the central engine and produce synchrotron photons. The polarized synchrotron photons travel alone the jet direction and then collide with the cold electrons at the front of the jet. After the Compton scattering process, these photons escape from the jet and are detected by the observer locating slightly off-axis. If photons are emitted before the bulk Lorentz factor saturates, the change of polarization angle is a natural result of the acceleration of the outflow.
\end{abstract}

\begin{keywords}
gamma-ray burst: individual (GRB 100826A) \--- polarization \--- radiation mechanism: non-thermal \--- scattering
\end{keywords}

\section{Introduction}\label{sec:introduction}

Gamma-ray bursts (GRBs) are the most energetic explosions in the universe since the Big Bang (for recent reviews, see, e.g.,  \citet{Piran:1999,Meszaros:2006}). They are divided into two categories (short and long) according to the duration $T_{90}$ -- the time interval in which 90\% of the photons are counted, $T_{90}>2$ seconds for long GRBs, and $T_{90}<2$ seconds for short GRBs\footnote{The threshold $T_{90}=2$ seconds is not strict, because in some cases a long lasting GRB seems to be a short one, and vice versa (see, e.g., \citet{Lu:2010}, and the references therein). Besides, a third group with intermediate duration can also exist (see, e.g., \citet{Ripa:2012} and the references therein).}. The isotropic equivalent energy radiated during the prompt phase of a GRB, which typically lasts a few seconds, is as large as the energy radiated from the sun during its whole life. After decades of extensive research since its discovery in the 1960s, it is still not clear how does the energy release in such a short time. The most accepted assumptions are the core collapse of a massive star \citep{Woosley:1993,Paczynski:1998,Fryer:1999}, or the merger of two compact objects, such as neutron star--neutron star or neutron star--black hole binaries \citep{Paczynski:1986,Goodman:1986,Eichler:1989,Meszaros:1997}. The light curves of GRBs have various types, ranging from smooth, fast-rise and quasi-exponential decay, through curves with several peaks, to highly variable curves with many peaks \citep{Meszaros:2006}. The spectra can often be fitted well by the Band function, and the $\nu F_{\nu}$ spectra generally peak at $0.1\sim 1$ MeV \citep{Band:1993}. Photons with energy higher than 1 GeV have also been detected in some brightest GRBs \citep{Abdo:2009a,Abdo:2009b,Abdo:2009c,Ackermann:2011}. In order to avoid the ``compactness problem", it is believed that gamma-rays are emitted from a highly relativistic outflow ejected by the central engine \citep{Rees:1966,Piran:1999,Ackermann:2010}. However, there are still controversies on some basic questions, e.g., is the outflow isotropic or collimated, magnetic-dominated or baryon-dominated? The information extracted from spectrum and light curve observation is not complete enough to distinguish one model from another.

The polarimetric measurement provides a deep insight into the nature of GRBs. Since the first detection of linear polarization in the optical afterglow of GRB 990123 \citep{Hjorth:1999}, linear polarization has been observed in many GRBs in the prompt phase \citep{Coburn:2003,Rutledge:2004,Wigger:2004,McGlynn:2007,Kalemci:2007,Yonetoku:2011,Yonetoku:2012,Berger:2011}, as well as in the afterglow \citep{Hjorth:1999,Covino:1999,Wijers:1999,Bersier:2003,Greiner:2003,Caldwell:2003,Steele:2009,Uehara:2012}. It seems that polarization in the prompt phase is often larger than that in the afterglow. The latter is usually less than $10\%$. The first detection of high linear polarization in the prompt phase was from GRB 021206, with polarization degree $\Pi=80\%\pm 20\%$ at a confidence level of $>5.7\sigma$ \citep{Coburn:2003}, although some independent groups could not conform this result \citep{Rutledge:2004,Wigger:2004}. \citet{Yonetoku:2011} reported a significant change of polarization angle between two adjacent time intervals in the prompt phase of GRB 100826A. They divided the most intense pulses of duration 100 seconds into two parts according to the shape of the light curve. The first part contains a single pulse lasting 47 seconds, while the second part contains multiple pulses lasting 53 seconds. They found that polarization angle changes $\sim 90^{\circ}$ between the two time intervals, while polarization degree almost keeps the same. A similar analysis to other two GRBs, GRB 110301A and GRB 110721A, however, shows no evident evolution of neither polarization degree nor polarization angle \citep{Yonetoku:2012}. \citet{McGlynn:2007} investigated the energy dependence of polarization in GRB 041219A, and found that polarization degree decreases as the photon energy increases, while polarization angle is almost independent of photon energy. \citet{Gotz:2009} also found the temporal variability of polarization degree and polarization angle in the prompt phase of GRB 041219A. As in the afterglow phase, the temporal evolution of polarization has also been observed \citep{Greiner:2003}. Note that the polarization mentioned here is referred to the linear polarization. The circular polarization is very small, especially in the afterglow phase \citep{Matsumiya:2003}. Besides, GRBs which have polarimetric observations all belong to the long category. The polarization of short GRBs is difficult to detect due to the short time of photon accumulation.

In the theoretical aspect, polarization may have different origins. It is well-known that synchrotron radiation can produce highly polarized photons. For the anisotropic and power-law ($N(E)dE\propto E^{-p}dE$) electrons, the polarization is known to be $\Pi_{\rm syn}=(p+1)/(p+7/3)$, if the magnetic field is globally uniform \citep{Rybicki:1979}. On the contrary, if the magnetic field is random on large scales, polarization is much reduced, or even vanishes. In the typical cases, the polarization induced by synchrotron radiation is expected to be less than $75\%$. The most prominent feature of polarization induced by synchrotron radiation is that it is energy-independent. Compton scattering is an alternative mechanism to produce high polarization. The polarization of an initially unpolarized photon scattered by a static electron, in the Thomson approximation, is $\Pi_{\rm comp}= (1-\cos^2\theta)/(1+\cos^2\theta)$, where $\theta$ is the scattering angle of the photon \citep{Rybicki:1979}. The polarization reaches its maximum at $\theta=90^{\circ}$. However, the probability of high polarization is small since the cross section reaches its minimum at this angle. Given some assumptions on the structure of the outflow, both synchrotron radiation and Compton scattering can produce a wide range of polarization \citep{Sari:1999,Gruzinov:1999a,Granot:2003,Lazzati:2004,Lazzati:2006,Toma:2009,Zhang:2011,Mao:2013gha}. Besides the above two intrinsic mechanisms, polarization can also originate from geometric effects \citep{Waxman:2003,Granot:2003}. If the jet open angle is small enough and the line-of-sight is close to the jet edge, the polarization signal is not completely averaged out. Thus, even randomly ordered magnetic field can produce high polarization. Nevertheless, all the above models can not naturally interpret the change of polarization angle observed in GRB 100826A. \citet{Yonetoku:2011} pointed out that the jet should be non-axisymmetric in order to cause the change of polarization angle. \citet{Lundman:2013qba} considered the polarization properties of photospheric emission originating from a highly relativistic jet, and found that significant degrees of linear polarization can be observed for the observers located at viewing angles larger than the jet opening angle. Particularly, the angle of polarization may shift by $\pi/2$ for the time-variable jets.

Synchrotron photons may suffer from the Compton scattering process in the optically thick region. Thus, both synchrotron radiation and Compton scattering process may contribute to the polarization. Recently, \citet{Chang:2013,Chang:2014a,Chang:2014b} investigated the polarization properties of the synchrotron-Compton process in the framework of the magnetic-dominated jet (MDJ) model. According to the MDJ model, the bulk Lorentz factor of the outflow increases as $\Gamma\propto r^{1/3}$, until reaching a saturated value $\Gamma_{\rm sat}$ \citep{Drenkhahn:2002a,Drenkhahn:2002b,Metzger:2010pp,Granot:2011,Meszaros:2011}. For typical long GRBs, $\Gamma_{\rm sat}\approx 250$, while it is much larger for short GRBs \citep{Chang:2012}. Electrons moving in the magnetic field radiate synchrotron photons. A beam of synchrotron photons with polarization degree $\Pi_0$ and polarization angle $\chi_0$ travels along the jet direction and then collide with cold electrons at the front of the jet. After the Compton scattering process, both polarization degree and polarization angle are changed. The polarization is expressed as a function of photon energy and viewing angle. The synchrotron-Compton model can produce a wide range of polarization, ranging from completely unpolarized, to completely polarized. Interestingly, at a special setup, the polarization angle can be changed $90^{\circ}$ exactly after scattering. We will show, in this paper, that the change of polarization angle observed in GRB 100826A can be naturally interpreted by the synchrotron-Compton model. \citet{Chang:2012} have shown that photons with energy $E \lesssim 100$ MeV are emitted before the bulk Lorentz factor saturates. If this is indeed the case, the change of polarization angle is a natural result of the acceleration of the outflow.

The rest of this paper is organized as follows. In section \ref{sec:model}, we briefly review the polarization of photons in the synchrotron-Compton process. In section \ref{sec:evolution}, we show the evolution of polarization angle with the bulk Lorentz factor. The $\sim 90^{\circ}$ change of polarization angle observed in GRB 100826A can be naturally interpreted as the acceleration of the outflow. Finally, discussions and conclusions are given in section \ref{sec:conclusion}.

\section{The polarization of photons in the synchrotron-Compton process}\label{sec:model}

Suppose a highly relativistic, magnetic-dominated and baryon-loaded (so electron-loaded) jet ejected from the central engine travels towards the observer with a large Lorentz factor. The magnetic field advected by the jet from the central engine is globally ordered \citep{Spruit:2001,Fendt:2004}. The jet consists of shells of slightly different velocities. When a fast shell catches up with a slow one, shocks are produced. The shocks accelerate electrons to power-law distribution, and at the same time magnify the magnetic field. The power-law electrons move in the magnetic field and radiate synchrotron photons. A beam of synchrotron photons, which are initially polarized, travel along the jet direction and then collide with the cold electrons at the front of the jet before escaping from it. The cold electrons here mean that the electrons are static with respect to the jet. After the Compton scattering process, both the polarization degree and polarization angle are changed. Since the jet is magnetic-dominated, the dynamical behavior of the jet seems as if it is consist of pure magnetic field, and the jet opening angle can be as small as $\sim 1/\Gamma$ \citep{Beloborodov:2010,Meszaros:2011}, where $\Gamma$ is the Lorentz factor of the jet.

Consider a polarized photon with energy $\varepsilon_0$ moves in the jet direction and then collides with a static electron. Define a Cartesian coordinate system such that the $z$-axis is along the jet direction, the $y$-axis is in the scattering plane, and the $xyz$ axes form the right-handed set. The energy of the photon after scattering is given as
\begin{equation}\label{eq:energy}
 \varepsilon_1= \frac{\varepsilon_0 }{1+ \frac{\varepsilon_0}{m_e c^2}(1- \cos\theta)},
\end{equation}
where $\theta$ is the scattering angle of the photon. The differential cross section for the scattering of a polarized photon by a static electron is written as \citep{Berest:1982,Chang:2014a,Chang:2014b}
\begin{equation}\label{eq:cross-section}
  d \sigma = \frac{1}{4} r_e^2 d \Omega \left(\frac{\varepsilon_1}{\varepsilon_0}\right)^2  \bigg[ F_0 +F_3(\xi_3+\xi'_3) + F_{11} \xi_1 \xi_1' +F_{22} \xi_2\xi'_2+F_{33} \xi_3\xi'_3\bigg],
\end{equation}
where $r_e=e^2/m_ec^2$ is the classical electron radius, $d\Omega=\sin \theta d \theta d \varphi$ is the infinitesimal solid angle, $\xi_i$ ($\xi'_i$) are the Stokes parameters of the incident (scattered) photon, and
\begin{equation} \label{eq:F}
\begin{cases}
 F_0=\displaystyle\frac{\varepsilon_1}{\varepsilon_0}+\frac{\varepsilon_0}{\varepsilon_1}-\sin^2 \theta,\\
 F_3=\sin^2 \theta,\\
 F_{11}=2 \cos \theta,\\
 F_{22}=\displaystyle\left(\frac{\varepsilon_1}{\varepsilon_0} +\frac{\varepsilon_0}{\varepsilon_1} \right) \cos \theta,\\
 F_{33}=1+ \cos^2 \theta.
\end{cases}
\end{equation}
Especially, if the incident and scattered photons are unpolarized, i.e., $\xi_i=\xi'_i=0$, equation (\ref{eq:cross-section}) reduces to the famous Klein-Nishina formula.

\begin{figure}
\centering
  \includegraphics[width=9 cm]{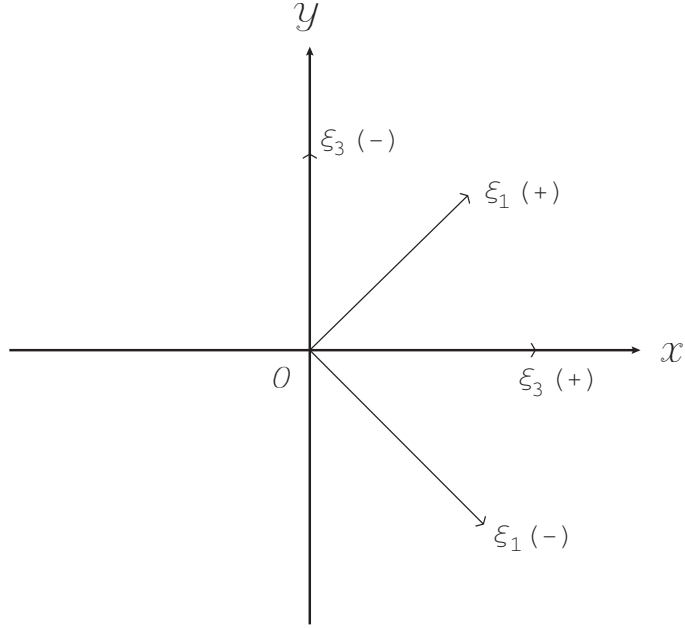}
  \caption{\small{The physical meaning of the Stokes parameters. The positive (negative) $\xi_3$ describes photons linearly polarized along the $x$-axis ($y$-axis). The positive (negative) $\xi_1$ means the linear polarization along the directions with azimuthal angles $+\pi/4$ ($-\pi/4$) relative to the $x$-axis in the $xy$ plane.}}\label{fig:stokes}
\end{figure}

The physical meaning of the Stokes parameters is presented in Fig.\ref{fig:stokes}.
The Stokes parameters stand for the polarization state of a photon. The positive (or negative) $\xi_3$ describes that the photon is linearly polarized along the $x$-axis (or $y$-axis). The positive (or negative) $\xi_1$ means the linear polarization along the direction with azimuthal angle $+\pi/4$ (or $-\pi/4$) relative to the $x$-axis in the $xy$ plane. The positive (or negative) $\xi_2$ stands for the right-handed (or left-handed) circular polarization. Since the circular polarization is very small in GRBs, we will ignore it in the following discussion. It should be paid specific attention that the Stokes parameters of the scattered photon are defined in a new coordinate system $O'x'y'z'$. The $O'x'y'z'$ system is the $Oxyz$ system rotating an angle $\theta$ relative to the $x$-axis, such that the $z'$-axis is along the moving direction of the scattered photon. This ensures that the polarization direction of the scattered photon is still perpendicular to its wave vector. The polarization degree of the incident photon is related to the three Stokes parameters by
\begin{equation}\label{eq:pi0}
  \Pi_0=\sqrt{\xi_1^2+\xi_2^2+\xi_3^2}.
\end{equation}
On the contrary, for a given photon with polarization degree $\Pi_0$ and polarization angle $\chi_0$, we can conveniently write its Stokes parameters as
\begin{equation}
\xi_1=\Pi_0\sin 2\chi_0,~~\xi_2=0,~~\xi_3=\Pi_0\cos 2\chi_0,
\end{equation}
where $\chi_0\in [-\pi/2,\pi/2]$ is the angle between the polarization direction and the $x$-axis.

Note that the Stokes parameters $\xi_i'$ are secondary quantities which essentially represent the properties of the detector as selecting one or the other polarization of the final photon, not the properties of the scattering process as such. The polarization states of the photon resulting from the scattering process itself are denoted by $\xi_i^{\rm f}$. They are given by the ratios of the coefficients of $\xi'_i$ in equation (\ref{eq:cross-section}) to the terms independent of $\xi'_i$ \citep{Berest:1982,Chang:2014a,Chang:2014b}, i.e.,
\begin{equation}\label{eq:xif}
\begin{cases}
  \displaystyle\xi^{\rm f}_1=\frac{ \xi_1 F_{11}}{F_0+\xi_3 F_3}=\frac{2 \Pi_0 \sin 2\chi_0 \cos \theta}{\varepsilon_1/\varepsilon_0+\varepsilon_0/\varepsilon_1-(1-\Pi_0 \cos 2 \chi_0)\sin^2 \theta},\\
  \displaystyle\xi^{\rm f}_2=\frac{ \xi_2 F_{22}}{F_0+\xi_3 F_3}=0,\\
  \displaystyle\xi^{\rm f}_3=\frac{F_3+ \xi_3 F_{33}}{F_0+\xi_3 F_3}=\frac{\sin^2 \theta+ \Pi_0 \cos 2\chi_0(1+\cos^2 \theta)}{\varepsilon_1/\varepsilon_0+\varepsilon_0/\varepsilon_1-(1-\Pi_0 \cos 2\chi_0)\sin^2 \theta}.
\end{cases}
\end{equation}
The circular polarization occurs only if the incident photon is circularly polarized ($\xi_2^{\rm f}\neq 0$ only if $\xi_2\neq 0$). After scattering, the polarization degree of the photon becomes to
\begin{equation}\label{eq:pi}
  \Pi=\sqrt{(\xi_1^{\rm f})^2+(\xi_2^{\rm f})^2+(\xi_3^{\rm f})^2},
\end{equation}
and the polarization angle is determined by
\begin{equation}\label{eq:chi}
 \tan2\chi=\frac{\xi_1^{\rm f}}{\xi_3^{\rm f}},
\end{equation}
where $\chi$ is the angle between the polarization direction and the $x'$-axis (which is the same to the $x$-axis). For any incident photon with polarization degree $\Pi_0$ and polarization angle $\chi_0$, we can derive the Stokes parameters of the scattered photon from equations (\ref{eq:energy}) and (\ref{eq:xif}). Then the polarization degree and polarization angle of the scattered photon can be further calculated from equations (\ref{eq:pi}) and (\ref{eq:chi}).

So far, we are working in the jet frame, which moves highly relativistically towards the observer. In order to transform to the observer frame, we note that the scattering angle between the two frames are related by \citep{Rybicki:1979}
\begin{equation}\label{eq:angle-transform}
  \cos\theta=\frac{\cos\bar{\theta}-\beta_{\rm jet}}{1-\beta_{\rm jet}\cos\bar{\theta}},
\end{equation}
where $\beta_{\rm jet}=(1-1/\Gamma^2)^{1/2}$ is the velocity of the jet in the unit of light speed, and $\Gamma$ is the bulk Lorentz factor of the jet. Here and after, symbols with a bar denote the quantities in the observer frame. In addition, the energy of the scattered photon in the jet frame can be Doppler-shifted to that in the observer frame, i.e.,
\begin{equation}\label{eq:doppler}
  \bar{\varepsilon}_1=\varepsilon_1\Gamma(1+\beta_{\rm jet}\cos\theta).
\end{equation}
Making use of equation (\ref{eq:energy}) and equations (\ref{eq:xif} -- \ref{eq:doppler}), one can express the polarization degree and polarization angle of the scattered photon as functions of $(\bar{\varepsilon}_1, \bar{\theta}, \Gamma, \Pi_0, \chi_0)$. In the next section, we will show how the polarization degree and polarization angle evolve with the bulk Lorentz factor of the jet.

\section{The change of polarization angle in the evolution}\label{sec:evolution}

\begin{figure}
\centering
  \includegraphics[width=10 cm]{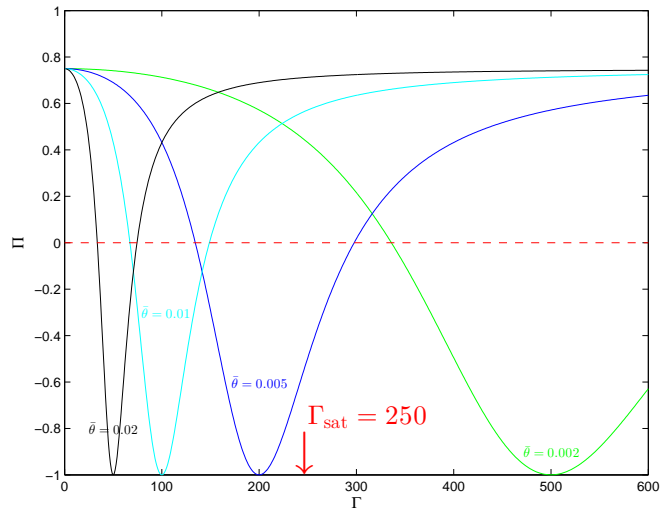}
  \caption{\small{Polarization degree as a function of the bulk Lorentz factor of the outflow at various viewing angles. The positive (negative) value of $\Pi$ means that the polarization direction is parallel (perpendicular) to the scattering plane.}}\label{fig:PI_075}
\end{figure}

For an observer near the earth, the viewing angle $\bar{\theta}$ is almost fixed. On the other hand, the polarimeters are designed to detect the polarization of photons in a certain energy band. Thus, the photon energy $\bar{\varepsilon}_1$ is a constant. Besides, $\Pi_0$ and $\chi_0$ do not evolve with time, since the photons produced by synchrotron radiation have unambiguous polarization degree and polarization angle. The only parameter that may evolve with time is the bulk Lorentz factor $\Gamma$. The MDJ model predicts that the outflow accelerates as $\Gamma\propto r^{1/3}$, until reaching a saturated value $\Gamma_{\rm sat}$. Then the outflow coasts with a constant velocity. In Fig.\ref{fig:PI_075}, we plot the polarization degree of the scattered photon, $\Pi$, as a function of $\Gamma$ for various viewing angles.
Fig.\ref{fig:PI_075} is a numerical result of equation (\ref{eq:pi}). In the numerical calculation, we set $\Pi_0=0.75$, which is the maximum polarization degree that synchrotron photons can achieve. We assume that the initial polarization direction is parallel to the scattering plane, i.e., $\chi_0=\pi/2$. For the photon energy, we set it to be $\bar{\varepsilon}_1=100$ keV. The $\nu F_{\nu}$ spectra of most GRBs peak at about $100-1000$ keV, and the polarimetric measurement are often carried out near this energy band. For example, the GAP on board IKAROS is fully designed to measure linear polarization in the prompt emission of GRBs in the energy band of $70-300$ keV \citep{Yonetoku:2006}. In Fig.\ref{fig:PI_075}, The positive and negative values of $\Pi$ mean that the polarization direction is parallel and perpendicular to the scattering plane, respectively.

From Fig.\ref{fig:PI_075}, we can see that the scattered photons can be completely polarized at the specific viewing angle $\bar{\theta}\Gamma\approx 1$ (corresponds to the valley of each curve). Most interestingly, the polarization angle can be changed $90^{\circ}$ at some special values of $\Gamma$ (corresponds to the intersection of each curve with the red-dashed horizonal line). This may provide a possible explanation for the change of polarization angle observed in the prompt phase of GRB 100812A \citep{Yonetoku:2011}. According to the MDJ model, the saturation bulk Lorentz factor of outflow for a typical long GRB is about $\Gamma_{\rm sat}\approx 250$, and low energy photons (say $\bar{\varepsilon}_1\lesssim 1$ MeV) are emitted before the bulk Lorentz factor saturates \citep{Chang:2012}. At a fixed viewing angle, e.g., $\bar{\theta}=0.005$, the polarization angle is initially parallel to the scattering plane. As $\Gamma$ increasing, the polarization degree gradually decreases to zero, and then increases again, but the polarization direction becomes to perpendicular to the scattering plane. Thus, the change of polarization angle is a natural result of the acceleration of the outflow. At a smaller viewing angle, say $\bar{\theta}\lesssim 0.002$, the acceleration of the outflow does not cause any change of polarization angle, although the polarization degree decreases. This may be the reason why no change of polarization angle was observed in the prompt phase of GRB 110301A and GRB 110721A \citep{Yonetoku:2012}. At a much larger viewing angle, say $\bar{\theta}\gtrsim 0.01$, the polarization angle may change two times. Initially, the polarization direction is parallel to the scattering plane. As $\Gamma$ increasing, the polarization direction changes to perpendicular to the scattering plane. When $\Gamma$ reaches a certain value, the polarization direction returns back to parallel to the scattering plane. However, this situation is not easily observed, since the flux reduces dramatically at large viewing angles.

\section{Discussions and conclusions}\label{sec:conclusion}

There are many explanations for the temporal evolution of polarization. \citet{Yonetoku:2012} pointed out that the change of polarization angle is due to the multiple patches of magnetic field whose characteristic angular size $\theta_p$ is much smaller than the jet opening angle $\theta_j$. If the jet open angle satisfies $\theta_j\sim \Gamma^{-1}$, we can see multiple patches with different magnetic field directions, and then significant change of polarization angle can be observed. On the contrary, if $\theta_j\gg \Gamma^{-1}$, we only see a limited range of the curved magnetic fields, thus no significant change of polarization angle occurs. \citet{Sari:1999} investigated the polarization and proper motion from a beamed GRB ejecta, and found that the polarization direction will change $90^{\circ}$ near the jet breaking time, if the offset of an observer from the center of the beam is large enough. The ICMART model proposed by \citet{Zhang:2011} also allows for the temporal evolution of polarization because the ordered magnetic field can be distorted by internal shock. In this paper, we have given an analytic calculation of the polarization properties of the synchrotron-Compton process in the framework of the MDJ model. We showed that both the polarization degree and polarization angle are changed after the Compton scattering process, regardless that the magnetic field is globally ordered and static. In the MDJ model, low energy photons are emitted before the bulk Lorentz factor saturates. As the jet accelerates, the change of polarization angle occurs naturally. If the initial polarization direction is parallel to the scattering plane, the polarization angle can change $90^{\circ}$ exactly. This provides a possible explanation for the polarization observed in the prompt phase of GRB 100826A.  Note that the change of polarization angle occurs only if the viewing angle is larger than $\sim \Gamma^{-1}$. Otherwise, if the viewing angle is much smaller than $\sim \Gamma^{-1}$, no change of polarization angle can be observed. GRB 110301A and GRB 110721A may belong to the later case.

The change of polarization angle is not necessarily to be $90^{\circ}$ exactly. It depends on the initial polarization angle $\chi_0$. If the initial polarization direction is neither parallel nor perpendicular to the scattering plane, the change of polarization angle can be any value between $0^{\circ} \sim 90^{\circ}$. The synchrotron-Compton model also allows for the variability of polarization. The high variability of light curve implies that the outflow may contain many shells with different Lorentz factors. Photons emitted from different shells may have different polarization degree and polarization angle. This is consistent with the temporal evolution of polarization observed in the prompt phase of GRB 041219A \citep{Gotz:2009}. In addition, the synchrotron-Compton model predicts that polarization degree of high energy photons is smaller than that of low energy photons \citep{Chang:2013,Chang:2014a,Chang:2014b}. Thus, the energy dependence of polarization detected in GRB 041219A can be naturally interpreted \citep{McGlynn:2007}. The recent analysis to GRB 061122 also shows a similar polarization-energy relation \citep{{Gotz:2013}}. The synchrotron-Compton model considered in this paper, although very simple, can account for many features of polarization of GRBs. Due to the low significance and limited amount of experiment data, it is still premature to exclude one model and favor another with the present data. More polarimetric measurement are needed to gain deep insight into the mechanism of GRBs. We hope that the future gamma-ray polarimeter POLAR\footnote{http://polar.ihep.ac.cn/cms/.} on board the Chinese Space Laboratory Tian-Gong II can provide more polarimetric data with unprecedented precision.

\section*{Acknowledgments}
We are grateful to X. Li, P. Wang, S. Wang and D. Zhao for useful discussion. This work has been funded by the National Natural Science Fund of China under Grant No. 11375203.

\label{lastpage}

\end{document}